\documentclass{article}
\usepackage{spconf,amsmath,graphicx}
\usepackage{booktabs}

\title{Conversational Co-Speech Gesture Generation via Modeling Dialog Intention, Emotion, and Context with Diffusion Models}
\name{
\vspace{-0.5em}
\begin{tabular}{@{}c@{}}
Haiwei Xue$^{1}$, Sicheng Yang$^{1}$, Zhensong Zhang$^{2}$, Zhiyong Wu$^{1,3,*}$,  \\ \textit{Minglei Li}$^{4, *}$, \textit{Zonghong Dai}$^{4}$, \textit{Helen Meng}$^{3}$ 
\thanks{* Corresponding author.}
\end{tabular}
}
\address{
    $^1$ Shenzhen International Graduate School, Tsinghua University, Shenzhen, China\\
    $^2$ Huawei Noah’s Ark Lab, Shenzhen, China\\
    $^3$ The Chinese University of Hong Kong, Hong Kong SAR, China\\   
    $^4$ Huawei Cloud Computing Technologies Co., Ltd, Shenzhen, China\\
    \small{
        xhw22$@$mails.tsinghua.edu.cn,
        yangsc21@mails.tsinghua.edu.cn,
        zhangzhensong@huawei.com,
    }\\
    \small{
        zywu$@$sz.tsinghua.edu.cn,
        liminglei29$@$huawei.com,
        daizonghong@huawei.com,
        hmmeng@se.cuhk.edu.hk
    }
}
\begin{document}
\ninept

\maketitle
\begin{abstract}
{
Audio-driven co-speech human gesture generation has made remarkable advancements recently. However, most previous works only focus on single person audio-driven gesture generation. We aim at solving the problem of conversational co-speech gesture generation that considers multiple participants in a conversation, which is a novel and challenging task due to the difficulty of simultaneously incorporating semantic information and other relevant features from both the primary speaker and the interlocutor. To this end, we propose CoDiffuseGesture, a diffusion model-based approach for speech-driven interaction gesture generation via modeling bilateral conversational intention, emotion, and semantic context. 
Our method synthesizes appropriate interactive, speech-matched, high-quality gestures for conversational motions through the intention perception module and emotion reasoning module at the sentence level by a pretrained language model. Experimental results demonstrate the promising performance of the proposed method.
}
\end{abstract}
\section{Introduction}
\label{sec:intro}

Co-speech gesture is very important in daily communication~\cite{1992Hand}, which complements the speech and makes the speech more engaging and vivid. 
For example,  ``you'' word is often accompanied by an implicit gesture of pointing the hand at the listener. When saying the ``cut" word, we may act the cutting gesture.
What's more, when someone laughs, his body will have a rhythmic tremor in most cases, and some people may hold their stomach.
In fact, there is an implicit connection between speech and gesture ~\cite{1970Movement}. 

Recently, audio driven co-speech gesture generation has drawn much attention from the community~\cite{DBLP:conf/cvpr/Yang00ZHBZ23, DBLP:conf/iccv/QianTZ0G21,DBLP:conf/ijcai/Yang0LZHBCX23,DBLP:journals/tog/YoonCLJLKL20,DBLP:journals/corr/abs-2305-18891}, due to its wide applications in the industry, such as digital human, CyberVerse, game and movie etc. 
Besides audio, some methods also consider other modalities, such as text and speaker identity information~\cite{DBLP:journals/tog/YoonCLJLKL20}. 

\begin{figure*}[t!]
    \centering
    \includegraphics[width=15.5cm]{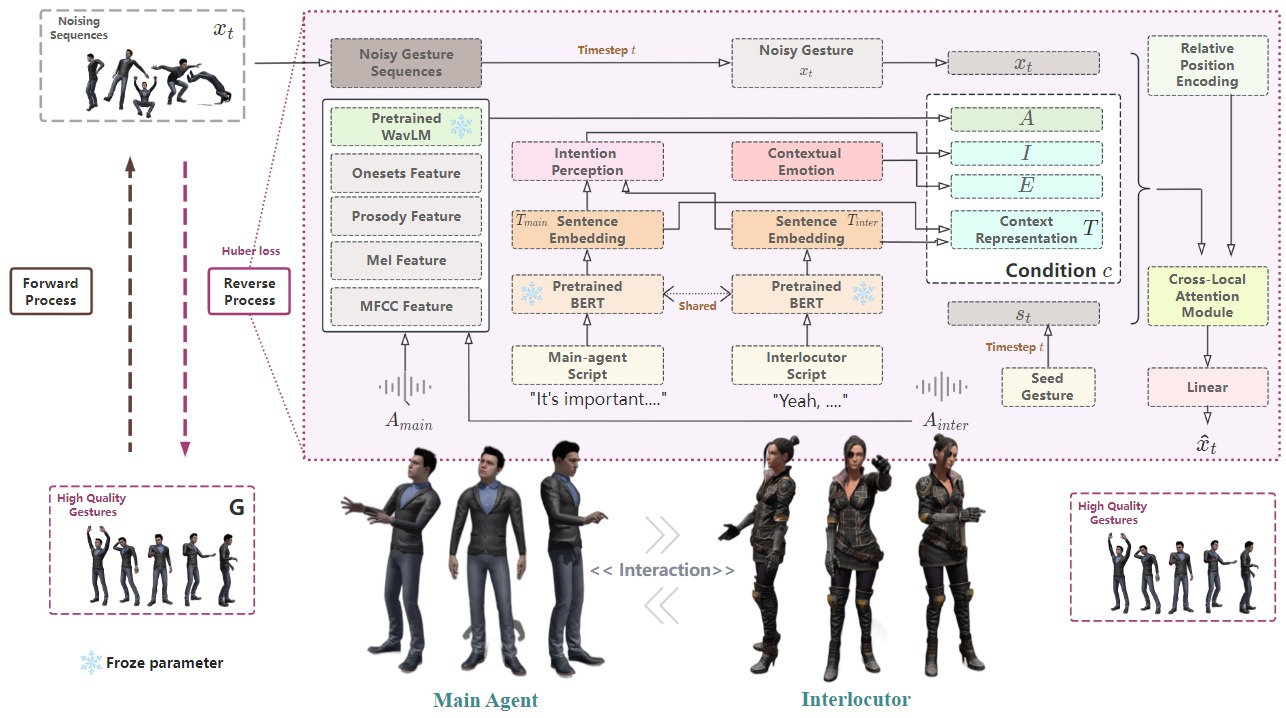}
    \caption{
    (Left) Forward process of CoDiffuseGesture is to add noise to the motion sequence from $t$ = 0 until $t$ = $T$. 
    (Right) Reverse process of CoDiffuseGesture is to learn the denoising ability. 
    A step $t$ and a noisy gesture sequence $X_t$ at the noising step conditioning on $c$ incorporating intent and emotion features $E, I$, audio $A$, seed gesture sequence $d$ are fed to the network.
    }
    \label{fig:diffusion}
\end{figure*}


While significant progress has been made, most of these works focus on single speaker speech-driven gesture generation. In real-life scenarios, conversations typically involve multiple persons, so speech-driven gesture generation in conversational interactions is under-studied. 
Yonatan Shafir et al.~\cite{DBLP:journals/corr/abs-2303-01418} is an earlier work that implements text driven gesture generation for two-person interaction with diffusion model.
Chopin et al.~\cite{DBLP:journals/corr/abs-2301-10134} introduces a bipartite graph approach to correlate multiple people's actions together to generate higher quality interaction actions.
But these works do not take into account the semantic information of speech, while Evonne Ng et al.~\cite{DBLP:conf/cvpr/NgJH0DKG22} considers two-person speech interaction, but they only model the face.

This paper aim at solving the novel task of multi-person conversational audio-driven gesture generation, which is challenging due to the following reasons: 
1) It is difficult for the main speaker's generated gesture to simultaneously encompass both the main speaker's and the interlocutor's speech semantic information; 
2) The gestures generated by the main speaker are struggling to appropriately accommodate the social characteristic of all participants in the conversation;
3) In multi-agent interactions, gesture generation lacks diversity and human-likeness.

To address the above challenges, we first employ a large pre-trained language model to infer sentence-level contextual representations instead of the word-level static word embedding approach in previous works.
Motivated by existing social psychology research~\cite{2019A,2019Emotional}, we observe that each person enters an interactive environment with specific emotions, attitudes, intentions, and behavioral tendencies. Based on this observation, we then introduce conversational intention and contextual affective information as social characteristics and leverage the fine-tuned language models for the two downstream tasks of intention and emotion. The inferred information is treated as a generative gesture condition.
Finally, we redesign the generation conditions and use a variant of the diffusion model to achieve more diverse and human-like gesture generation for conversational interaction.

The main contributions of our paper are: 
(1) We model contextual information and conversational intent, allowing the generated gestures to have more appropriate and human-like expressions across various conversational topics.
(2) We propose to consider the emotional aspects of all participants, aiming to associate the speaker's emotions with their corresponding actions.
(3) We are the first to apply the diffusion model to the multimodal conversational speech-driven interaction gesture generation task. We propose an interaction gesture generation framework via encoding speech semantics, dialog intents, and emotions of multiple participants to generate higher quality gesture motions through the diffusion model. 

\section{methodology}
\label{sec:methodology}

\subsection{Problem Definition}

We aim to synthesize the high-quality gestures of the main speaker with the multimodal information of the dialogue as input, including the main speaker's speech and script, the interlocutor's speech and script, the social feature of the dyadic non-verbal interaction.
Given two speakers' (the main agent and the interlocutor) audio and their corresponding speech scripts as inputs, the goal of our model $F$ is to infer continuous diverse appropriate, human-like co-speech gestures as $G\in[G_1, .... , G_N]$, where $N$ denotes the length of speech audio $A$ and script $T$. Here, $G_i$ represents $J$ joints of the human body including two hands, which can be visualized in Blender.

Inspired by the Dyadic Nonverbal Interaction model~\cite{2019A}, when two people have a conversation, the intention of the conversation and the emotion of the speech have a significant impact on the gestures of the speaker.
Therefore, to obtain diverse co-speech gestures that are more human-like and appropriate for agent speech and interlocutor behavior, we introduce a conversational intention label $I$ and a multi-label emotion intensity value $E$ to provide supervision during the training phase.
Formally, given a conversational sequence clip, there is a generation condition $c = [A, T, I, E]$ for synthesizing main agent gestures.
Intent features are extracted from a fine-tuned pre-trained model of Distibert-Base-Uncased~\cite{Sanh2019DistilBERTAD} trained on a massive dataset. Notice that the intent label actually is the one-hot vector in our approach.
The emotion label $E$ contains positive, neutral, and negative emotion intensity values.
In our approach, we employ a learning framework and procedure that resembles MDM~\cite{DBLP:journals/corr/abs-2209-14916}.
The forward process adds Gaussian noise to the gesture sequence at each diffusion step $t$. 
The reverse process is trained by optimizing the Huber loss in given condition $c$. The architecture of our proposed CoDiffuseGesture is shown in Figure~\ref{fig:diffusion}.

\subsection{Modeling Dialog Intention, Emotion, and Context}

Inspired by Dyadic Nonverbal Interaction Systems (DNI)~\cite{2019A} and Basic Emotion Theory (BET)~\cite{2019Emotional}, we observe that each person enters an interactive environment with specific emotions, attitudes, intentions, and behavioral tendencies.
Therefore, we model the intention, emotion, and context information of dyadic non-verbal interactions in generation condition $c$.
The DNI model introduces a mental model in dyadic non-verbal interactions: 1) Perceptual processes ; 2) Cognitive resources + cognitive-affective; 3) Processes; 4) Goal; 5) Behavior; 6) Interaction.
Through social psychological model~\cite{barrett2011context}, it is found that one of the important factors affecting the interaction and the resulting behavior is the goal, that is, the intention of the two parties during the conversation.
This is also consistent with our intuition that when a conversation involves asking for directions, the person giving directions often combines gestures to indicate the correct path. And during casual conversations, there may be more body shakes caused by laughter.

Hence, modeling the intention during the conversation can make the generated gestures more appropriate to the interaction content of the current conversation.
We use a fine-tuned version of Distilbert-Base-Uncased~\cite{Sanh2019DistilBERTAD} on the massive dataset to reason the possible conversation intentions coarse-grainedly.  The massive dataset includes nearly 60 different conversational intention labels. Such as transportation, music preferences, interest chat, daily socializing, takeout, etc.
Based on the DNI architecture, the intention of the two parties to chat is not invariable, and they may change during the interaction process. So the intention is continuously inferred from the short timestamp. It should be noted that the result of the final inference will be mapped to the one-hot vectors with the dimension of 60.

Another important influence on interactive gestures is emotion, which BET theory~\cite{2019Emotional} states is fundamentally about instigating action and changing the probabilities of future actions.
Some studies~\cite{scarantino2017things, fischer2017theory} on emotion also point out that emotional expression conveys four kinds of information about interaction: (1) how the speaker is feeling at the moment; (2) what is happening in the current context; (3) the perceived action or behavior expected by the other person; and (4) the intention or plan that the two parties may make.
The most common example is when a person is extremely angry and is likely to make fist gestures. Thus, the emotions of both parties in a conversation are one of the most important influence factors, which is confirmed in the cognitive-affective processes in the DNI architecture.

Based on the above research~\cite{2019Emotional, DBLP:journals/corr/abs-2305-18891}, we decided to introduce the emotional information of both parties in the generation of interactive co-speech gestures.
We employed the Roberta~\cite{liu2019roberta} model, which has been pre-trained by Twitter data, to extract the emotional information of both parties at the sentence level.
Afterward, we deduce three kinds of emotional intensity of positive, neutral and negative, symboling emotional information $E$.
Finally, to enhance the understanding of semantic content, we utilize a pre-trained language model~\cite{devlin2018bert} to represent contextual semantic information at the sentence level instead of FastText~\cite{bojanowski2017enriching} in previous work~\cite{DBLP:conf/icmi/ChangZK22}. Finally, the generative condition $c$ consists of speech information $A_{main} ~A_{inter}$, dialogue intention features $I$, emotion features of both parties $E$, and contextual dialogue semantics $T_{main}~T_{inter}$.

\subsection{Diffusion Training Architecture}
In our work, we follow the DiffuseStyleGesture~\cite{DBLP:conf/ijcai/Yang0LZHBCX23} framework, but instead of restricting to a single agent only, we generalize to two-person dyadic interactions for conversational co-speech gesture generation.
Similar to most diffusion model methods, both include the forward process $F_{+}$ and the reverse process $F_{-}$ for training.

{\bf Forward Process} The forward process involves repeatedly adding a small amount of Gaussian noise to the real data until the data exhibits Gaussian noise. Formally, the forward process on a real sample from a real data distribution consists of a Markov chain with gradually increasing noise.  The distribution will eventually resemble a standard Gaussian.
\begin{equation}
    x_t, ~t = F_+(G, \mathcal{N}(0, I))
\end{equation}

{\bf Reverse Process} The reverse process is also called the denoising module. This module needs to be trained together under $x_t$ pure noise, the noise steps $t$, the seed gesture $s_t$, and the conditions $c$. We predict the signal $\hat{x}$ itself rather than the noise in each noise step in denoising process.
\begin{equation}
    \hat{x} = F_-(x_t, s_t, t, c)
\end{equation}
where $c=[A,T,I,E]$. The audio features $A$, context $T$, and emotion intensity $E$ contain multimodal representations of the two individuals in a conversation, which are concatenated modal features of the main agent and the interlocutor, respectively.
To train the reverse process for denoising, we optimize the Huber Loss between the generated and the real gestures.
After the training is completed, in the inference procedure, it is only necessary to process the speech and text into corresponding features as conditions $c$ in the same way.
And then randomly sample part of the gesture sequence from the training data as the seed gestures $s$. 
Then pure noisy $x_t$, the seed gestures $s$, step $t$, and the condition $c$ can be feed into the trained denoising module to get the final predicted gesture result.
Modeling the interaction information of two people can improve the quality of gesture generation, according to the experiment.
\section{experiments}
\label{sec:experiment}

\subsection{Dataset}
Currently, there is still a shortage of high-quality datasets for conversational co-speech gesture generation. Fortunately, the GENEA2023 dataset~\cite{DBLP:journals/corr/abs-2308-12646} provides approximately 20 hours of excellent dyadic non-verbal interaction co-speech gesture data.
This dataset is derived from the original Talking With Hands (TWD)~\cite{DBLP:conf/iccv/LeeDMSSS19} data, which is processed by the organizer to provide both sides' voice, speaker ID, and both sides' motion information saved through BVH motion capture files. Conversation topics are mostly related to our daily interactions.
Each annotated clip at millisecond accuracy contains essentially only 1–2 words. Meanwhile, laughter is specially labeled, which is represented by the ``\#'' symbol.
The GENEA 2023 dataset contains a total of 483 dialogs including audio files, script and BVH motion capture files of both the main agent and the interlocutor.
There are 372 dialogs in the training set, 41 in the validation data, and 70 in the test data. The total amount of the data is around 20 hours. The test data lacks the gesture file of the main agent speaker, which needs to be predicted by the model.
In particular, 29 of the 70 test set of data do not match the main speaker and the interlocutor data, which can be used for the study of matching interlocutor consciousness.
\begin{figure*}[t!]
    \centering
    \includegraphics[width=14.5cm]{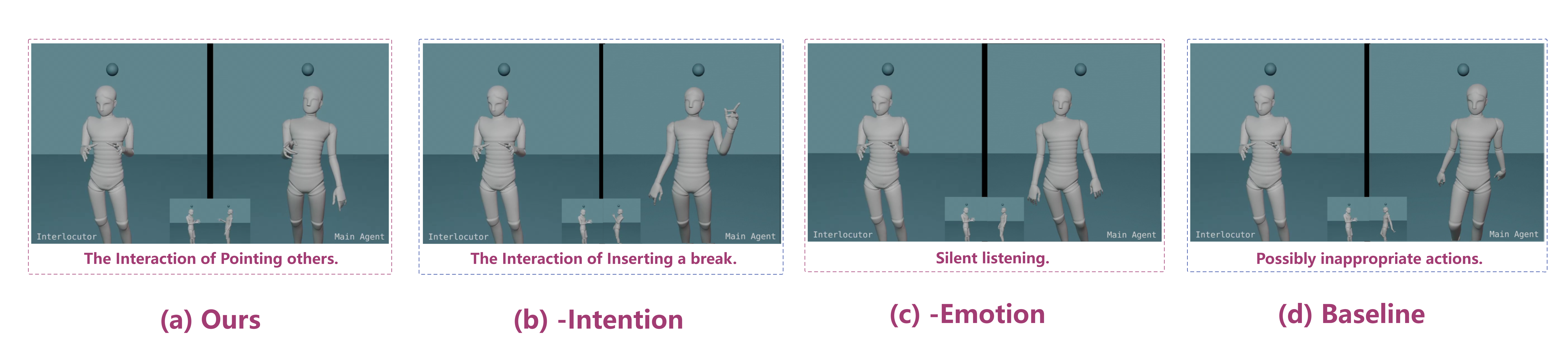}
    \caption{
    Visualization of different gestures reflected by the four variational models.   
    }
    \label{fig:viz}
\end{figure*}
\subsection{Experiment Setup}

We merge the training and validation data and split each data into 5-second clips of 30 FPS meta-training data for updating the model. 
The first second is treated as a seed gesture, and the subsequent duration is used to minimize the loss between the real and generated gestures to train the network.
All feature is normalized before feeding into the model to reduce the risk of gradient explosion or dispersion.

We train our model on an NVIDIA A100 40G GPU with a batch size of 400 and AdamW optimizer with learning rate is $3* 10^{-5}$.  We train about a week on GENEA2023 for 500,000 epochs with $T$ = 1000 noising steps and a cosine noise schedule.
We evaluate the quality of dyadic nonverbal interaction gesture generation from three aspects: human-likeness, appropriateness for agent speech, appropriateness for interlocutor status.

During the experiment, participants score three aspects while watching gestures generated by blender renderings visualizations.
The participants primarily consisted of graduate students and working professionsals, ranging in age from 20 to 40.
Each evaluation aspect contains 1–5 points, which are excellent, good, fair, poor, and bad. 
We ensured that the participants remained unaware of the specific method variant associated with each video during the rating process.
For the human-likeness assessment dimension, we asked the participant the question, ``Does he resemble human movements? Do his actions resemble the kind of gestures a man might make?”
We evaluate the appropriateness of main agent speech by asking the participants whether the main agent's gesture matches his speech and has a certain sense of rhythm.
For the final evaluation indicator, we asked participants the question, ``When the interlocutor is talking, what gesture might the main agent make? Is the listener's response human-like?”
Based on the results report, it is found that our proposed method has good quality for generating dyadic nonverbal interaction co-speech gestures in conversation.
It can take into account both speaking and listening states to make appropriate and human-like gestures.

\subsection{Experimental Results and Analysis}

\begin{table}
\centering
\caption{Evaluation results of different variant model. 
Human-likeness, Appropriateness for Agent Speech (AAS), and Appropriateness for Interlocutor status (AIS) are results of MOS with 95\% confidence intervals. `-' is shorthand for `without' in ablation studies. `*' denotes the proposed method.
}
\label{jieguotu}
\vspace{0.1cm}
\resizebox{\columnwidth}{!}{%
\begin{tabular}{@{}lccc@{}}
\hline
              & Human-likeness & AAS & AIS \\ \midrule
Baseline      & 3.625$\pm$0.1816          & 3.663$\pm$0.1573                           & 3.494$\pm$0.1883                                 \\
Ours*          & \textbf{3.875$\pm$0.1481}          & 3.775$\pm$0.1298                            & 3.606$\pm$0.1681                                 \\ 
~~- Intention & 3.763$\pm$0.1536         & \textbf{3.831$\pm$0.1510}                          & 3.550$\pm$.1517                                    \\
~~- Emotion   & 3.781$\pm$0.1515        & 3.813$\pm$0.1366                           & \textbf{3.650$\pm$0.1487}                                    \\

\hline
\end{tabular}
\vspace{-0.5cm}
}
\vspace{-0.8cm}
\end{table}

Table \ref{jieguotu} shows the evaluation results of different variant models rated by invited participants.
Each metric samples nearly 200 video ratings from nearly 30 participants, and the higher the score, the better the model performs.

{\bf Quantitative Results} In the comparison of existing model experiments, the baseline was chosen to win second place ~\cite{DBLP:journals/corr/abs-2308-13879} in the GENEA 2023 competition for comparison experiments and is easy for us to compare due to its open-source reproducibility.
From the table above, our proposed method achieves a relatively good effect on human-likeness, with a score of 3.875 $\pm$ 0.1481 in the 95\% confidence interval, which is significantly different from the baseline of 3.625 $\pm$ 0.1816.
It is obvious that appropriateness for Agent Speech (AAS) and appropriateness for Interlocutor status (AIS) have improvement in comparison to the baseline.

{\bf Ablation Study} We conduct ablation studies to demonstrate the effectiveness of each of the proposed components, as shown in the bottom of Table \ref{jieguotu}.
The ablation studies utilize the symbol ``-" as a shorthand for without.
The experimental results shown in the table that the variation of without intention information has achieved the best effect among all experiments in the Appropriateness for Agent Speech (AAS), with a subjective score of 3.831 $\pm$ 0.1510, while only add the emotional information of the dialogue between the two parties to the conditions.
This may indicate that emotional ups and downs have a certain implicit correlation with changes in gestures, for example, when happy, it may be accompanied by body and hand shaking, which vaguely reveals the happy mood of the speaker.
On the other hand, under the variant model without emotional information, the matrix Appropriateness for Interlocutor status (AIS), which involves the interlocutor, achieved the best performance with a score of 3.650 $\pm$ 0.1487 when the intention of dialogue between the two sides was taken into account.
This is also very consistent with the characteristics of sociology, and the intention of dialogue between the two sides has a relatively important impact on the gestures of the two sides, so it may obtain relatively good results in the AIS matrix.
Compared with the three indicators, it can be clearly seen that the AIS associated with the information of both sides has a low score, indicating that the current model needs to improve the modeling ability of the information of both sides.

{\bf Visualization and Discussion } 
We visualized some gesture prediction results using the blender with python script from~\cite{DBLP:conf/icmi/YoonWKVNTH22}, as shown in Figure \ref{fig:viz}.We also strongly recommend the reader to visit our demo video\footnote{https://youtu.be/JHkyoI0qFNA} for better visualization.
The first panel (a) of the figure shows the proposed approach, where an interaction is initiated from the main agent by pointing to the interlocutor for giving responses when the interlocutor is talking about himself, which is in line with our daily habits.
In the second panel (b), the emotion-related variants also interact with each other over a wider range of movements while expressing content, as opposed to remaining quiet and even developing some joint problems in the two right panels (c-d).
In the baseline, there is a more obvious hand back bend, which is not in line with our human habits.

\section{conclusion}
\label{sec:conclusion}

Most previous works have focused on single-person gesture generation. In this paper, inspired by the work on social dyadic interactions, we propose a diffusion-based architecture to model the semantics, conversational intent, and emotions of both parties for high-quality, human-like, and appropriate conversational co-speech gesture generation.
The results of the experiments demonstrate that taking into account the semantics of speech, intentions, and emotions of both participants in a conversational scene under the generation condition can enhance the system's performance, especially the method based on the diffusion model, capable of generating diverse and human-like gestures.
While promising, there are still many questions and further research to be investigated.
For instance, only the sociological characteristics of intention and emotion have been considered in our work. In fact, social knowledge, gender, and the surrounding environment have the potential to influence the gestures of both parties during their communication.
Moreover, regarding design generation conditions, we are in the coarse-grained phase of global information extraction. It is possible to explore the application of named entity recognition to the semantics of a dialogue between two parties.
These interesting problems are worth further research in the future.

\section*{Acknowledgments}

This work is supported by National Natural Science Foundation of China (62076144), National Social Science Foundation of China (13\&ZD189), Shenzhen Key Laboratory of next generation interactive media innovative technology (ZDSYS20210623092001004) and Shenzhen Science and Technology Program (WDZC202208161 40515001, JCYJ20220818101014030).

\bibliographystyle{IEEEbib}
\clearpage
\bibliography{main}
\end{document}